%% file: auto-kernel-sel.tex
\documentclass[final,5p]{elsarticle}
\usepackage{graphicx}
\usepackage[caption=false]{subfig}
\usepackage{booktabs}

\makeatletter
\def\ps@pprintTitle{%
 \let\@oddhead\@empty
 \let\@evenhead\@empty
 \def\@oddfoot{\hfill\thepage\hfill}%
 \let\@evenfoot\@oddfoot}
\makeatother

\begin{document}

\begin{frontmatter}

\title{%
  Performance portability through machine learning guided kernel selection in SYCL libraries
}
\author{John Lawson}
\address{Codeplay Software Ltd.}
\ead{john@codeplay.com}

\begin{abstract}

  Automatically tuning parallel compute kernels allows libraries and frameworks
  to achieve performance on a wide range of hardware, however these techniques
  are typically focused on finding optimal kernel parameters for particular
  input sizes and parameters.  General purpose compute libraries must be able to
  cater to all inputs and parameters provided by a user, and so these techniques
  are of limited use.  Additionally parallel programming frameworks such as SYCL
  require that the kernels be deployed in a binary format embedded within the
  library. As such it is impractical to deploy a large number of possible kernel
  configurations without inflating the library size.

  Machine learning methods can be used to mitigate against both of these
  problems and provide performance for general purpose routines with a limited
  number of kernel configurations. We show that unsupervised clustering methods
  can be used to select a subset of the possible kernels that should be deployed
  and that simple classification methods can be trained to select from these
  kernels at runtime to give good performance. As these techniques are fully
  automated, relying only on benchmark data, the tuning process for new hardware
  or problems does not require any developer effort or expertise.

\end{abstract}

\begin{keyword}
  Auto-tuning\sep SYCL\sep GPGPU\sep Machine learning\sep Performance
  portability
\end{keyword}

\end{frontmatter}

\input{introduction}
\input{matmul}
\input{pruning}

\input{classification}

\input{vgg}
\input{conclusions}

\section*{Acknowledgements}
The author would like to thank Duncan McBain and Daniel Soutar for thoughtful
comments and interesting discussions about this work. This research did not
receive any specific grant from funding agencies in the public, commercial,
or not-for-profit sectors.

\section*{References}
\bibliographystyle{elsarticle-num}
\bibliography{bibliography}

\end{document}

%% file: introduction.tex
\section{Introduction}\label{sec:intro}

Auto-tuning has been widely studied as a technique to allow libraries to obtain
portable performance across a range of devices by utilising parameterized
kernels and selecting the right parameters to match the compute capabilities of
the different devices.

For frameworks like OpenCL that provide their kernels as source
code this works especially well. The source code can be configured using the
preprocessor to handle any number of possible parameter configurations. Other
parallel programming frameworks like CUDA and SYCL provide the kernels in a
compiled binary format, and so each set of parameters requires a new binary blob
containing the kernel compiled with those parameters.
Supporting many different kernel instantiations in these libraries adds
complexity and a cost in terms of library size and build times.

Standard auto-tuning techniques sample the kernel parameter space in order to
determine the set of parameters that give the
best performance for a given problem. This process aims to provide the absolute
best performance for that particular set of input sizes and problem parameters
and so is especially effective when these inputs and problem parameters are
constant. On the other hand the auto-tuning must be done every time the inputs
or parameters change, which is typically a costly process.

As a result of this it is difficult to use auto-tuning to provide general
purpose libraries that can cater to all possible inputs. We look at using
unsupervised machine learning techniques to explore the space of kernel
parameters and select a subset of kernels that can be deployed in a library to
provide close to optimal performance on a wide range of possible inputs. These
clustering techniques allow the library to achieve over 90\% of the optimal
performance while limiting the library to include as few as four kernels.

We also consider how well machine learning classification methods can select
from these kernels at runtime. Decision trees are an effective way to do this,
preserving a large proportion of the possible performance while being easy to
integrate into the library.

When combined, these automated approaches are an effective way to extract
performance from parameterized kernels suited for a wide range of possible
inputs, and this performance can be achieved with very little developer effort.
These approaches allow a simple matrix multiplication kernel to provide
performance similar to or even much better than hand optimized BLAS
implementations on a range of hardware. We demonstrate this by comparing the
inference time of VGG16, a popular image classification network implemented
using SYCL-DNN, an accelerated neural network library, when using different
matrix multiplication routines. The tuned simple kernel is competitive on
desktop GPUs and performs better than optimized BLAS libraries on integrated
GPUs and mobile GPUs.

\section{Background and related work}

\subsection{SYCL and OpenCL}\label{sec:sycl}

OpenCL~\cite{opencl} is a heterogeneous programming framework developed
originally by Apple and now maintained by the Khronos Group. It is an open
standard designed to provide a cross platform way to program a wide range of
hardware from GPUs to FPGAs.  OpenCL allows developers to write compute kernels
in a subset of C, which are embedded within applications and libraries as
strings of source code. This source code is then just-in-time (JIT) compiled to
match the target device at runtime.

By using JIT compilation, OpenCL allows developers to use the preprocessor to
inject constants and types into generic kernels. Different versions of the same
kernels can be compiled multiple times to match the different inputs and sizes
at runtime while using the same source code. As the same source code can be used
for all the different parameter values, there is no cost to using this technique
beyond the additional compilation time to compile each kernel.

SYCL~\cite{sycl} is a more recent open standard from the Khronos Group,
introduced in 2014 aiming to remove the boilerplate and complexity of lower
level heterogeneous programming frameworks like OpenCL. Using SYCL a developer
can write compute kernels using standard C++ as well as make use of the strong
C++ type system to track data dependencies and manage data movement between host
and device.

OpenCL requires hardware vendors to package a C compiler with their device
drivers, but to support SYCL it would be more
challenging to include a full C++ compiler and JIT compile heavily templated
C++ kernels. Instead, SYCL adopts a two stage compilation approach, where the
kernels are initially compiled to an intermediate representation (IR) that is bundled
with the library or application binary. This IR blob is then passed to the
OpenCL JIT compiler at runtime, significantly reducing the amount of work
required to compile the kernels at runtime.

The downside of shipping kernels in a binary format is that these now include
the kernel parameters, and so a different binary blob is required for each
instantiation of the kernel.

There are many existing OpenCL and SYCL accelerated compute libraries,
including the BLAS implementations clBLAS~\cite{clblas},
CLBlast~\cite{nugteren2018clblast} and
SYCL-BLAS~\cite{syclblas,aliaga2017sycl}. Each of these libraries is tuned for
their target hardware to some extent. These libraries either provide a set of
hardcoded kernel parameters for given inputs chosen by hand to try to give good
performance, or include more automated approaches that include benchmark
scripts that generate these sets of parameters that can then be compiled into
the library. These automated approaches currently use heuristics and limited
numbers of kernel benchmarks to try to establish which parameters to use.

\subsection{Auto tuning}

There are many auto-tuning techniques that have been widely studied. General
purpose tuning frameworks such as clTune~\cite{cltune} and Kernel
Tuner~\cite{kernel_tuner} provide easy to use tuning for compute kernels, tuning
OpenCL, CUDA and other kernels. The techniques used by these frameworks combine
kernel benchmarks to measure the performance of a given of parameters, and a
parameter search algorithm to selectively sample from the parameter space while
maximising performance.

Despite the sophistication of these parameter search algorithms, such
auto-tuning systems can be expensive in terms of power and time usage, and must
be run for each required set of inputs. This can be partially mitigated using
machine learning to learn a model of the kernel performance and using this model
to predict reasonable parameters to start the auto-tuning search. Techniques
discussed in~\cite{ml_autotuning} and in~\cite{boosted_trees_tuning} replace the
parameter search algorithms with machine learning based approaches. A random
sample of kernel configurations are benchmarked, and these timings used to train
a model that predicts the timings of all other kernel configurations, allowing
the optimal configuration to be directly chosen from the predicted times.

Other uses of machine learning in automated kernel optimization include
predicting whether an operation would be computed faster on CPU or
GPU~\cite{prog_mapping_a, prog_mapping_b}, and whether a kernel would perform
better when manually caching data in local memory~\cite{ml_local_mem}.

Auto-tuning has been used to provide portable performance on different hardware
for a variety of different computational tasks, including
convolutions~\cite{tuning_convs}, matrix
multiplication~\cite{note_tuning_gemm,nugteren2018clblast},
 FFTs~\cite{tuning_ffts} and
stencils~\cite{autotuning_stencils,autotuning_3d_stencils}.

A different approach to auto-tuning is to explore the different kernel
parameters during the end program runtime. This dynamic approach of auto-tuning
allows the best available configuration to be found if the same problem is
computed multiple times. This is used in the TensorFlow~\cite{tensorflow} and
MXNet~\cite{mxnet} machine learning frameworks with the cuDNN~\cite{cudnn}
launcher options. While this does not provide as fine grained control as the
kernel based auto-tuning techniques, it does allow coarse grained decisions
about the best algorithm or approach to take for given problems on fixed
hardware.

%% file: matmul.tex
\section{A matrix multiply case study}

Matrix multiplications are an integral part of modern deep learning and many
other domains, so having accelerated routines optimized for particular hardware
gives a significant impact on the performance of these computations. The kernels
that calculate a matrix multiplication have been the target of many previous
auto-tuning techniques as the kernels can easily be written to make use of many
parameters. These kernels are less complicated compared to other stencil or
convolutional kernels, while having enough scope for loop transformations,
tiling and caching memory accesses that they are good targets for tuning.

In~\cite{ashes_paper} we introduced a matrix multiply case study using the
parameterized kernels provided by the SYCL-DNN~\cite{sycldnn} library. This
paper continues the study of auto-tuning these kernels, expanding the number of
benchmarks and the devices targeted by the tuning techniques.

Each work item in this matrix multiplication kernel computes a small tile of the
output. For integers $R, A, C$, it loads an $R\times A$ tile from the left hand
input and an $A \times C$ tile from the right hand input, which are accumulated
into an $R \times C$ output tile. These tile sizes are compile time constants
that also correspond to the vector sizes used to load the values from memory, so
the possible values are 1, 2, 4 and 8. These three parameters give 64 different
kernel configurations.

In addition to the compile time kernel constants we considered the effects of
different work group sizes on performance, using a combination of 1, 8, 16, 32,
64 and 128. As the total work group size for a kernel is limited by the device
drivers, we only used the following pairings: (1, 64), (1, 128), (8, 8),
(8, 16), (8, 32), (16, 8), (16, 16), (32, 8), (64, 1) and (128, 1); giving a total
of 640 possible configurations to select from.

To measure the effects of the different kernel parameters and work groups sizes
we ran a number of benchmarks on two platforms. With only 640 possible
configurations it is feasible to test the performance of every configuration.
This allows us to evaluate whether the kernel selection techniques manage to
choose the best performing kernel and avoids any confounding factors that may
arise when combining these techniques with standard kernel auto-tuning
techniques. As auto-tuning will typically try to selectively search the kernel
parameter space it will only end up sampling the performance of some kernel
configurations and so would immediately discount some kernels from being chosen.

Fully connected and convolutional layers in machine learning models can be
computed using matrix multiplications. The SYCL-DNN library is designed to
provide accelerated routines for machine learning models so matrix sizes derived
in this way are representative of the typical workloads for the library. The
benchmarks use the matrix sizes from three popular neural networks:
VGG~\cite{vgg}, ResNet~\cite{resnet} and MobileNet~\cite{mobilenetv2}. Overall
these gave 300 different sets of sizes for the input matrices of the
computations.

\subsection{Data collection}

The benchmarking framework used to collect the data ran a small number of warmup
iterations to ensure the devices were running at optimal clock speeds and that
the kernels were compiled. The measurement collected was the total time for a
number of iterations of kernel execution, giving an overall mean time for each
kernel execution.  The actual number of iterations varied depending on the time
of execution, aiming for each benchmark to run for around 1 second in total.
Between each benchmark run the framework paused for a short amount of time to
help reduce any thermal throttling, and device temperatures were monitored
during the benchmarking process to ensure there was no throttling.

The devices used to run the benchmarks were:
\begin{itemize}
  \item An AMD R9 Nano GPU (driver v2482.3).
    \item An Intel i7-6700K CPU (driver v18.1.0.0920).
\end{itemize}
We used SYCL on top of OpenCL to target these devices, providing the kernels as
SPIR.

\subsection{The dataset}

The matrix sizes in the dataset vary, with some being very large and others
small, some fairly square with a large batch size and others very tall and
skinny. These different sizes provide different performance characteristics for
the kernels on the hardware. For example the tall and skinny matrices lead to
very few threads being used in the multiplication and so for large compute
devices like the AMD GPU a lot of the compute capacity goes unused. Even using
auto-tuning to select the best kernel will not solve this problem, and really a
separate kernel should be used that is designed to utilize all the hardware for
these sorts of matrix inputs. This is beyond the scope of the paper, but should
such a kernel be available then the type of kernel could be considered as
another parameter that has to be selected by a tuning system.

\begin{figure}
  \centering
  \includegraphics{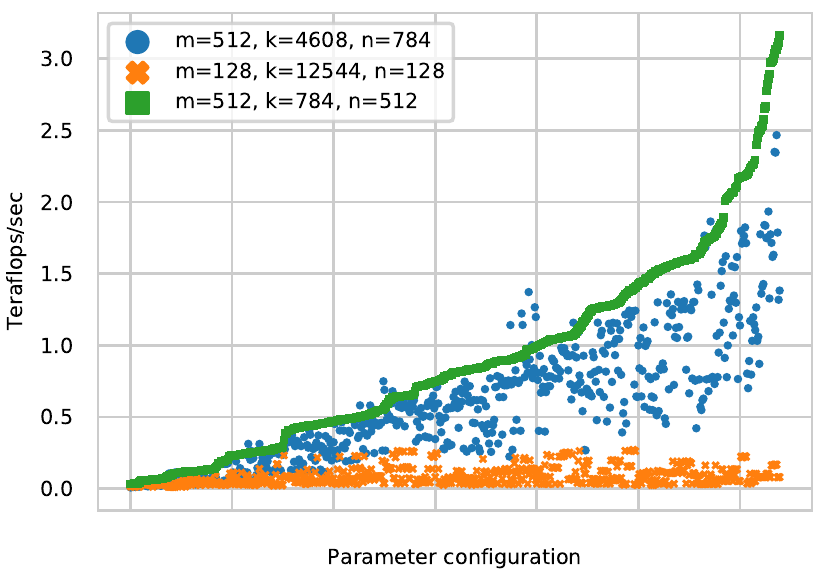}
  \caption{The performance of all different kernel configurations for three
  sets of input sizes on the AMD R9 Nano GPU, with varying matrix sizes from
  square to rectangular. Multiplying small reasonably square
  matrices performs best overall and favors large tile sizes, while tall skinny
  matrices perform poorly in all configurations.}
  \label{fig:sample}
\end{figure}

As an example of this, on the R9 Nano the best performing configuration (tiles
(8, 4, 4), work-group (16, 16) for m=512, k=784, n=512, batch=16) achieves  3160
gigaflops per second, while the worst configuration (tiles (1, 8, 1), work-group
(8, 8) for m=32, k=12321, n=27, batch=1) only achieves 13 Gflops/sec. The best
configurations for the small cases are the ones that use the most threads and so
achieve the highest utilisation of the GPU, while the best configurations for large
problems are the ones that reuse the most data without spilling registers. As
the numbers of threads and numbers of registers are device specific, these are
the things that an automated kernel deployment system would have to implicitly
learn from the dataset.

Figure~\ref{fig:sample} shows the performance for three different sets of input
matrix sizes. The more square matrices (m=512, k=784, n=512) allowed the kernels
to perform best, but it only achieved optimal performance in a very small number
of kernel configurations. In this case, of the 640 possible configurations only
55 achieved over 2 teraflops/sec and only 7 of those achieved over 3
teraflops/sec.  This highlights the importance of tuning the kernel parameters
and ensuring that the best parameters are available in a library.

The second results in Figure~\ref{fig:sample} from a more rectangular set of input
matrices (m=512, k=4608, n=784) have three kernel configurations that achieve
over 2 teraflops/sec. All three of these kernel configurations achieve over 3
teraflops/sec with the square input sizes, but the best performing configuration
for the square inputs achieves less than 1.4 teraflops/sec for the rectangular input
sizes. The third set of results correspond to an input set with a very large
number of elements to accumulate, and as discussed above the kernel used is not
optimized for these cases and so performs poorly overall.

\begin{figure}
  \centering
  \subfloat[AMD R9 Nano]{%
    \includegraphics{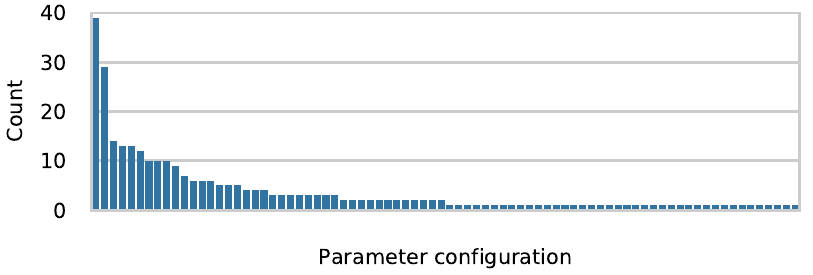}%
  }

  \subfloat[Intel i7-6700K CPU]{%
    \includegraphics{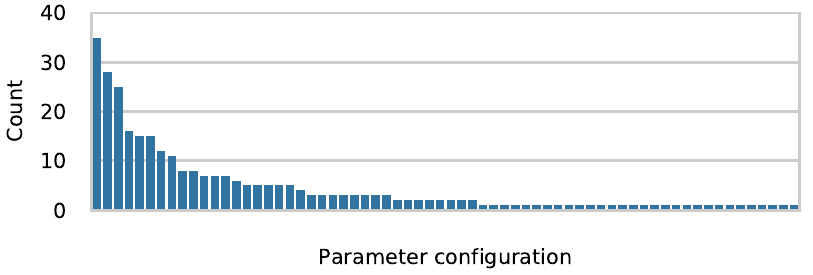}%
  }

  \caption{The number of times a configuration of kernel parameters achieves
  optimal performance in the dataset. For the AMD GPU, one configuration is best
  in 39 cases, but 80 distinct configurations are best in at least one case. For
  the Intel CPU the top three configurations are best in 35, 28 and 25 cases
  respectively and 68 are best in at least one case.}%

  \label{fig:config_counts}%
\end{figure}

The challenge faced by an automated kernel selection program is that many different
configurations obtain the best performance for different matrix sizes.
Figure~\ref{fig:config_counts} shows that while there are a small number of
configurations that perform best in a large number of cases, there is a long
tail where many other configurations also perform best in at least one of the
benchmarks. This long tail illustrates the problem with pruning the number of
configurations required to deploy within a library, and suggests that any such
pruning will result in some loss of performance. The goal of this paper is to
determine whether an automated solution can minimize this loss in performance.

The dataset and the corresponding code is
available online~\cite{jwlawson_kernel_tuner}. The machine learning routines
were provided by scikit-learn~\cite{scikit-learn}.

\subsection{Determining the target number of configurations}

As discussed in Section~\ref{sec:sycl} a SYCL library cannot deploy an unlimited
number of kernels, as they are embedded within the library as binary blobs. As
such the kernels that should be deployed must be carefully selected to provide
as much performance as possible.  The number of kernels to deploy could be
determined through trial and error by investigating the achievable performance
of different numbers of kernels. A more tractable approach would be to explore
the variance within the dataset and use that to estimate how many kernels may
encapsulate that variance.

\begin{figure}
  \centering
  \subfloat[AMD R9 Nano]{%
    \includegraphics{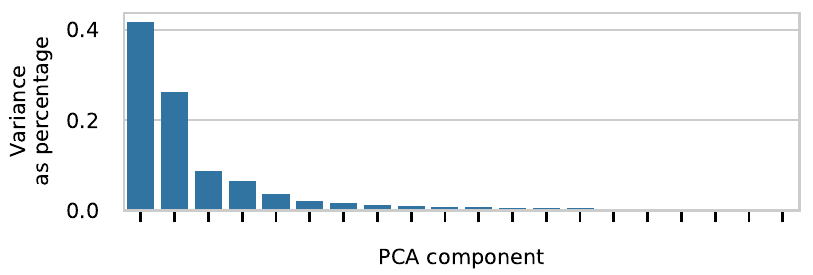}%
  }

  \subfloat[Intel i7-6700K CPU]{%
    \includegraphics{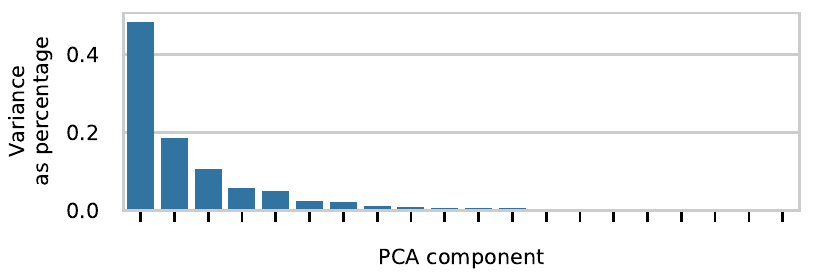}%
  }
  \caption{The percentage of the variance of the dataset accounted for by
  each PCA component. For the AMD GPU over 80\% of the variance is accounted for in
  the 4 main components, 90\% is accounted for in 7 components, and 95\% in 14.
  For the Intel CPU 4 components account for 80\% of the variance, 6 components
  for 90\% and 11 components for 95\%.}%
  \label{fig:pca_variance}%
\end{figure}

Principal component analysis (PCA)~\cite{pca_original,prob_pca} finds a new
coordinate system for the dataset that concentrates the variance into as few
dimensions as possible. In this way these principal dimensions contain the most
distinguishing information about the dataset. Figure~\ref{fig:pca_variance}
shows the amount of total variance in the dataset that is accounted for by each
of the components identified by PCA. This highlights that the data is fairly
structured and that the majority of the variance is encapsulated within a small
number of components.

As PCA shows that most of the dataset's variance can be encapsulated in less
than 15 components we study how much performance can be encapsulated when
providing at most 15 kernels. We compare the performance that is achievable when
the number of kernels that would be deployed in a library varies between 4 and
15.

\subsection{Normalization method}\label{sec:norm}

For each set of matrix sizes, the benchmarks measured the performance as
gigaflops per second for each kernel. This gives 640 floating point values
describing the performance, ranging from 0 to the maximum Gflops/sec of the
device.

When comparing the performance of kernels for fixed matrix sizes, it is helpful
to consider the comparative performance of the kernels instead of the raw
flops/s achieved. By normalizing the data to only show the comparative
performance, the data is easier for an automated system to understand.  Such a
normalization technique should map the performance to a value between 0 and 1,
with the best performing kernels valued at or close to 1, while poor performing
kernels have a value closer to 0.

In the original work, the only normalization technique considered was to scale
the performance results relative to the performance of the kernel that performed
best. The normalized value is obtained by dividing the achieved performance by
the maximal performance for a fixed input. This provides a uniform mapping that
preserves the relative performance between all kernels.

As the kernel selection process should infer more from the better performing
kernels than the worst performing kernels, and hopefully never tries to select
kernels that give mediocre performance, we can normalize the data to only
preserve the kernels that perform well. We study three different approaches of
doing this.

The first approach is to use a raw cutoff point, so that all results under a
certain threshold are clamped to 0. In the results below we consider a cutoff
value at 90\% of the peak performance, so all results that obtain less than 90\%
of the optimal performance for each set of inputs is set to 0. This introduces
sparsity in the data but does not change any non-zero values, so they range
between 0.9 and 1.

An extension of this is to rescale the normalized data after clamping the poorly
performing kernels. This ensures that the values make full use of the 0 to 1
range but may encourage the models to discard good performing kernels that it
thinks actually perform poorly. In the discussion below we refer to this as the
standard cutoff normalization technique (as opposed to the raw cutoff).

A final approach studied is to use a modified sigmoid function to map the scaled
values, with many of the less well performing kernels mapped to 0. The sigmoid
function $f(x) = (1 + \exp(50 * (0.85 - x)))^{-1}$ was constructed to map 85\%
performance to 0.5 with all values less than 80\% mapped to less than 0.1.

\begin{figure}
  \centering
  \includegraphics{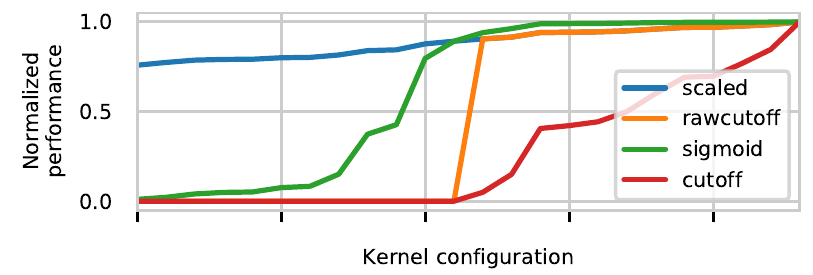}
  \caption{Comparison of different data normalization techniques for the best
  performing set of input sizes for the AMD GPU.}
  \label{fig:norms}
\end{figure}

Figure~\ref{fig:norms} shows the effects these normalization techniques have on
the best performing set of inputs for the AMD GPU, with the raw performance
shown in Figure~\ref{fig:sample}. As the normalization techniques all
clamp low performing kernel configurations to zero, only the configurations
achieving over 75\% of the performance of the best configuration are shown.

%% file: pruning.tex
\section{Kernel selection}\label{sec:pruning}

The techniques in this paper to deploy kernels in SYCL libraries is made up of
two steps. First the kernel configurations should be selected, and then a simple
model is constructed to choose which of these configurations to use at runtime
for a given problem. As SYCL kernels are embedded into the library as binaries
it is impractical to include a large number of kernels. In order to balance
performance and binary size the number of kernels must be pruned to those that
give the best performance on a range of different problems.

The initial selection of kernels is done using unsupervised clustering of the
dataset. For a given set of matrix sizes the dataset provides performance
information for each of the 640 kernel configurations. This performance
information can be represented as a point in 640-dimensional space, though as
the raw times vary between matrix sizes it is useful to normalize these
coordinates.

Matrix sizes that have similar performance characteristics will naturally end up
with similar coordinates, and so clustering techniques can be used to group
these together. By considering these clusters of similarly performing matrix
sizes we can extract which kernels give the best performance.

\subsection{Clustering techniques}\label{sec:clustering}

There are many unsupervised machine learning clustering techniques available
which try to extract meaning directly from the data. These each have different
behaviors and consider different aspects of the data, so may extract widely
varying sets of kernels.

\subsubsection{$K$-means clustering}
A relatively simple clustering method is $k$-means clustering, which is an
iterative method to find $k$ centroids that minimize the distance from each
points in the dataset to their closest centroid. This method is effective when
the clusters have shapes that are close to the unit ball in the coordinate
space, however if the cluster shapes are less regular or intertwined the method
will struggle to separate the clusters.

\subsubsection{PCA and $k$-means clustering}
To help get around this, the coordinate space of the dataset can be transformed
to help separate the clusters. One approach to do this is using Principal
Component Analysis to reduce the dimensionality of the dataset and concentrate
the variance of the dataset by making use of the full range of values in each
of the new dimensions, then using $k$-means clustering on this transformed data.

\subsubsection{Spectral clustering}
Another similar approach is to use a spectral transformation before using
$k$-means clustering. A similarity graph of the coordinates in the dataset can
be represented as an adjacency matrix. The eigenvectors of the Laplacian of this
matrix provide new coordinates that can be clustered using $k$-means.

\subsubsection{HDBScan}
Density based methods can also be used to cluster data, which use the
density of the data to establish the boundaries between clusters.
HDBScan~\cite{hdbscan_paper,hdbscan_software} is an example of such a clustering
method that uses a hierachical tree structure to construct the clusters and
provide better estimates of outlying data.

Unlike the other clustering methods, HDBScan does not provide a parameter for
the number of target clusters, rather providing however many clusters it finds
based on its other hyperparameters. In order to limit the numbers of clusters we
compute the numbers of clusters for a sweep of the hyperparameters and in the
following use whichever values gave the correct number of clusters.

\subsubsection{Decision tree}
While not a clustering method, decision trees can be used to choose a subset of
a dataset by artificially limiting the number of leaf nodes in the tree. A
decision tree can be trained as a regression solver that maps the input matrix
sizes to the vector of performance data. Unlike the clustering methods, this
takes into account the matrix sizes rather than just the performance data. Each
leaf node then ends up being a performance vector which is an approximate
representative of the performance vector for all input sizes that end up at that
node in the tree.

\subsection{Selecting the kernels to deploy}



To compare the effectiveness of clustering methods for selecting kernel
configurations to deploy in a library we explored their outcomes given the
benchmark dataset.  We used a selection method of choosing the kernels that gave
best performance by count. This Top-N method is a formulation of the methods
used when previously selecting the kernels manually, and serves as a useful
baseline to see how differently more advanced methods perform.

The clustering methods provide either representatives of the clusters, such as
the centroids of the $k$-means clusters, or just the cluster labels for each of
the data entries. When there are representatives of the clusters, these can be
used to select an optimal kernel by looking at which kernel configuration
performs best for the representative. When the full cluster is provided the
optimal kernel is computed by taking the geometric mean of all elements in the
cluster and choosing the best performing configuration of this mean set of
values.

\subsection{The results}

The dataset was split into training and test subsets, allowing a comparison of
how well the techniques generalize to previously unseen matrix sizes. Each
proposed technique used the training dataset to select a fixed number of kernel
configurations and the test dataset was used to evaluate what percentage of the
optimal performance could be achieved by only considering those selected
kernels.

The optimal performance of the test data is given by the benchmark data and
normalized to between 0 and 1. A geometric mean of each value for the best
performing kernel out of the selection was computed with all entries of the test
dataset to give this final performance figure.

\begin{figure}
  \centering
  \includegraphics{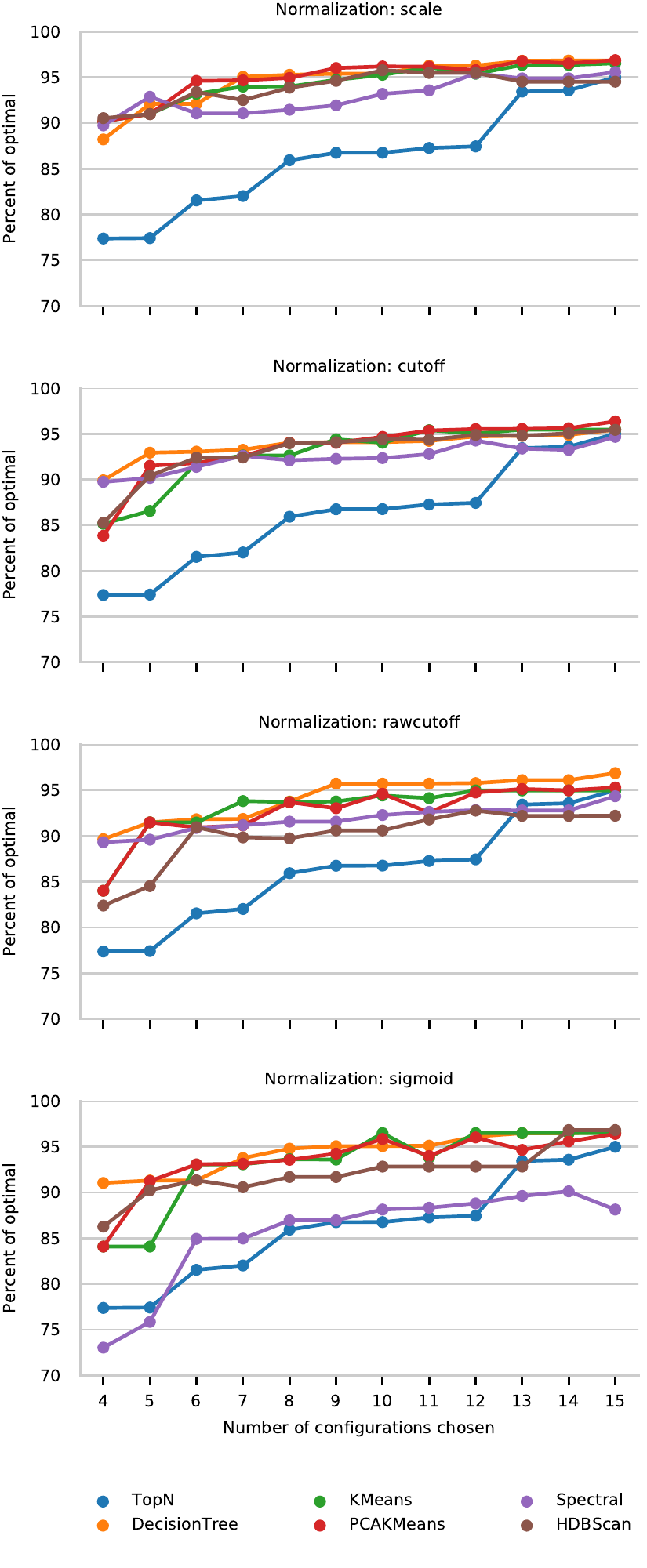}
  \caption{The performance of each pruning technique in
  Section~\ref{sec:pruning} as a percentage of the optimal obtainable
  performance for the AMD R9 Nano GPU, comparing the normalization techniques
  discussed in Section~\ref{sec:norm}.}%
  \label{fig:amd_error_classes}%
\end{figure}

\begin{figure}
  \centering
  \includegraphics{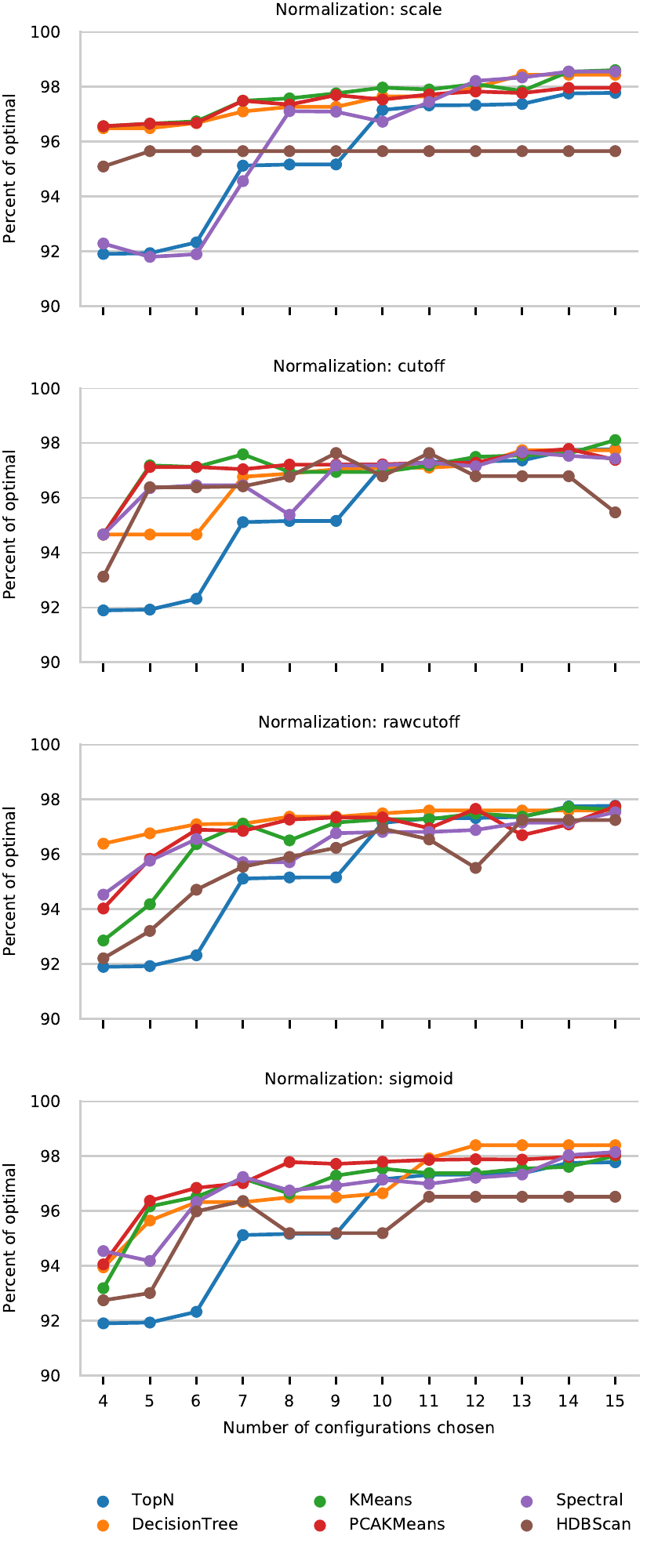}
  \caption{The performance of each pruning technique in
  Section~\ref{sec:pruning} as a percentage of the optimal obtainable
  performance for the Intel i7-6700K CPU, comparing the normalization techniques
  discussed in Section~\ref{sec:norm}.}%
  \label{fig:intelcpu_error_classes}%
\end{figure}

Figure~\ref{fig:amd_error_classes} shows the percentage of the optimal
performance obtained by the different clustering techniques on the AMD GPU for
the four different normalization techniques discussed in Section~\ref{sec:norm}.
The machine learning methods all perform better than the Top-N method of
selecting the kernels based on those that perform best by count, except when the
number of kernels selected gets very large. Some of the selection methods
perform almost as well when selecting as few as 6 kernels, and don't improve
much as the number of kernels increases. This suggests that there are a small
number of kernels that perform well for a wide range of input sizes, but that
are not the ones that actually perform best for a large number of inputs.

For example when the number of kernels is limited to 4, the 4 top kernels by
count are:
\begin{itemize}
  \item Tiles (4, 8, 4),  work-group (16, 16)
  \item Tiles (4, 8, 4),  work-group (8, 16)
  \item Tiles (4, 8, 4),  work-group (8, 32)
  \item Tiles (8, 4, 4),  work-group (8, 32)
\end{itemize}
The tile sizes are all similar, with slightly different work-group sizes. These
configurations perform similarly, and must perform well for some of the most
common input sizes. However they do not perform well on the large number of less
optimal input sizes, and so overall this selection gives poor performance. In
comparison the decision tree selection is:
\begin{itemize}
  \item Tile (2, 8, 1), work-group (8, 32)
  \item Tile (2, 8, 4), work-group (16, 16)
  \item Tile (4, 4, 4), work-group (8, 32)
  \item Tile (4, 8, 4), work-group (8, 32)
\end{itemize}
It includes only one of the top performing configurations, but this allows the
overall kernel selection to be better suited to the different corner cases.
These much more varied configurations therefore give better performance across a
wider range of the input sizes.

All clustering methods performed well for the standard scaled normalization,
though the Spectral clustering method performed worst after TopN. For the more
sparse normalization techniques the performance of the clustering methods start
to become more varied. Both the decision tree and k-means methods appear to
perform well across the different normalization techniques, while the
performance of HDBScan can vary.

This is promising for extending this data to the much more sparse data that
would be generated by other auto-tuning techniques that run benchmarks of many
fewer configurations. In these cases the data will naturally be much more sparse
than the brute force dataset, and these normalization techniques mimic the data
that might be obtained from these approaches.

The clustering methods most affected by normalization method are HDBScan and
spectral clustering. When the data becomes more sparse these methods appear to
select less optimal kernels and therefore gain worse performance overall. In
addition HDBScan was the hardest to train, as the numbers of clusters cannot be
specified as a parameter, so a parameter search is required to select the best
options to limit the numbers of kernels.

Figure~\ref{fig:intelcpu_error_classes} shows the same data but for the Intel
i7-6700K CPU. In the benchmarks this device was more consistent in the
performance that it achieved for different input sizes. As such all kernel
selection techniques performed significantly better than for the AMD GPU, where
there was much more variation in the obtained performance.

In these benchmarks, the HDBScan density based clustering technique performed
surprisingly poorly and the results varied significantly depending on the
number of kernels. For the standard normalization technique all tested
parameters gave only 4 or 5 kernels.

The decision tree clustering method performed well for the AMD data, often
achieving among the best performance, however for the CPU this is not the case.
It seemed to lose the least performance on the raw cutoff normalization scheme,
but for all other normalization schemes the decision tree tends to be outperformed by
the other clustering methods.

\subsection{Clustering conclusions}

The baseline option of choosing the kernels by which appear to be best most
often is a weak approach. The more intelligent clustering methods outperformed
this in the majority of cases, as they consider the distribution of the data
more generally and use that to select the kernels that provide better
performance across a wide range of inputs.

The aim for the kernel clustering is to automatically prune the number of
kernels to provide in a library. As such the chosen solution should provide good
performance regardless of the device or normalization scheme.
The decision tree, spectral clustering and HDBScan clustering give varied
performance across the devices and types of normalization, whereas the K-means
and PCA+$K$-means clustering methods provide stable and good results. There are
definitely cases where these relatively simplistic clustering techniques do not
perform as well as some others, but the difference is rarely large.

%% file: classification.tex
\section{Deploying the kernels}\label{sec:classification}

Selecting which kernels to deploy in a library is only half the story as our
goal is to be able to support any inputs required by our users. This requires a
method to map the user's inputs to the best kernel configuration provided by the
library. Such a process must be carried out before launching each kernel to
ensure that the optimal choice is made at each point. This means that the
selection process must be both effective and inexpensive to compute; there is
little point gaining a small performance boost in the kernel if it is outweighed
by time spent in a large classification system.

\subsection{Classifiers}

The previous sections investigated how to limit the number of kernel
configurations that should be provided in a library. Selecting which of these
kernels to run is a classification problem that maps the input matrix sizes to
the optimal kernel configuration. For each entry in our dataset we can see which
of the chosen kernels provides the best performance, and train a classifier to
do this selection using standard supervised learning techniques.

There are many different techniques for classification using machine learning.
The classifier will have to be run each time a new matrix multiplication is
launched by the library and so the main challenge is to balance the
effectiveness of the classifier with the time taken to make a classification.
More complicated state of the art classifiers like neural networks may be very
effective, but they are also computationally expensive and so would be a poor
choice to integrate in this way.  Decision trees on the other hand are easy to
implement in a performant way and easy to integrate in a library, as they can be
implemented as a series of nested if statements within the kernel launcher. If a
decision tree can effectively infer the best kernel to use for unseen matrix
sizes then this would be an ideal solution to use.

To establish whether this is the case, we compare the effectiveness of three
decision trees to other classification techniques. The decision trees have
increasing limits on the depth and numbers of samples allowed for leaf nodes.
Varying these parameters helps establish how much the decision tree might be
overfitting.  Deeper trees can fit better to the training data, but will
potentially overfit to suit the training data and perform poorly on previously
unseen inputs.

The three decision trees are signified A, B and C. Decision tree A has no limit
on the maximum depth and allows splitting down to single sample leaf nodes if
required. Decision tree B has a maximum depth of 6 and requires leaf nodes to
have at least 3 samples, while decision tree C has a maximum depth of 3 and
requires at least 4 samples at the leaves. There are many other possible
combinations of parameters, however additional tuning of these risk overfitting
to the testing data set.

\begin{table}
  \caption{The performance results for the classifiers as a percentage of the
  absolute optimal performance, for the kernel configurations selected by
  PCA+$K$-means for the AMD R9 Nano. Note that the maximum achievable
  performance for the selection of configurations is limited to 91.19\%,
  94.62\%, 94.94\% and 96.89\% for the 5, 6, 8 and 15 configurations
  respectively.}
  \label{tab:amd_class}
  \centering
  \begin{tabular}{@{}rcccc@{}}
  \toprule
    & \multicolumn{4}{c}{\textbf{Number of configurations}} \\
  \cmidrule(l){2-5}
    \textbf{Classifier} & \textbf{5} & \textbf{6} & \textbf{8} & \textbf{15} \\
  \midrule
       DecisionTreeA & 88.16 & 86.82 & 85.53 & 85.64 \\
       DecisionTreeB & 86.10 & 90.62 & 83.21 & 83.01 \\
       DecisionTreeC & 84.56 & 85.39 & 82.30 & 83.66 \\
    1NearestNeighbor & 77.37 & 78.93 & 77.79 & 75.48 \\
    3NearestNeighbor & 78.15 & 78.64 & 76.85 & 76.82 \\
    7NearestNeighbor & 75.38 & 74.85 & 75.08 & 77.39 \\
           LinearSVM & 68.68 & 74.46 & 67.31 & 77.62 \\
           RadialSVM & 70.93 & 70.93 & 70.93 & 70.93 \\
        RandomForest & 86.91 & 89.31 & 87.60 & 83.96 \\
                 MLP & 63.61 & 56.35 & 64.39 & 62.99 \\
  \bottomrule
  \end{tabular}
\end{table}

\begin{table}
  \caption{The performance results for the classifiers as a percentage of the
  absolute optimal performance, for the kernel configurations selected by
  PCA+$K$-means for the Intel i7-6700K CPU. Note that the maximum achievable
  performance for the selection of configurations is limited to 96.55\%,
  96.65\%, 97.34\% and 97.95\% for the 5, 6, 8 and 15 configurations
  respectively.}
  \label{tab:intelcpu_class}
  \centering
  \begin{tabular}{@{}rcccc@{}}
  \toprule
    & \multicolumn{4}{c}{\textbf{Number of configurations}} \\
  \cmidrule(l){2-5}
    \textbf{Classifier} & \textbf{5} & \textbf{6} & \textbf{8} & \textbf{15} \\
  \midrule
       DecisionTreeA & 91.65 & 92.59 & 93.50 & 92.29 \\
       DecisionTreeB & 93.14 & 91.86 & 93.87 & 90.15 \\
       DecisionTreeC & 92.26 & 91.11 & 91.51 & 91.28 \\
    1NearestNeighbor & 91.36 & 91.36 & 91.40 & 89.73 \\
    3NearestNeighbor & 91.18 & 90.26 & 91.61 & 86.42 \\
    7NearestNeighbor & 88.00 & 90.15 & 89.22 & 87.96 \\
           LinearSVM & 84.18 & 76.20 & 88.32 & 85.64 \\
           RadialSVM & 80.49 & 83.80 & 78.55 & 83.80 \\
        RandomForest & 93.65 & 93.90 & 93.26 & 93.85 \\
                 MLP & 74.30 & 79.23 & 79.23 & 76.88 \\
  \bottomrule
  \end{tabular}
\end{table}

Nearest neighbor is another relatively simple classification technique that
classifies an input based on which of the training inputs are closest to it. As
such it requires that the training dataset be stored alongside the classifier to
compute which data points are the input's neighbors. As such it would be
infeasible to deploy within the library but provides a useful comparison for
what similar classifiers can achieve.

Other classifiers are more complex and require significantly more computation to
infer a class from an input. Classifiers like SVM, which computes the vectors
that separate the classes, and random forest ensembles, made up of multiple
decision trees that are combined together, can potentially provide better
performance but would require more work on the host when choosing the kernel to
launch.

The comparisons made between these classifiers considered how well they could
infer the optimal kernel given the subset of kernels provided by the pruning
techniques discussed in Section~\ref{sec:pruning}. As the choice of kernels is
limited to this subset the maximum achievable performance is not 100\%.

Tables~\ref{tab:amd_class} and~\ref{tab:intelcpu_class} show the relative
performance of the different classification methods for a range of possible
kernel configurations. Overall the decision tree classification methods perform
well, in many cases significantly better than the more computationally expensive
methods.

One of the more surprising observations here is that the performance does not
improve as the number of classes does, despite the theoretical maximum
achievable performance increasing. The absolute best performance for both
devices was obtained with just 6 kernel configurations, and the decision
tree obtaining best performance for either 6 or 8 kernel configurations. While
the additional kernel choices may allow higher theoretical performance, the
models seem to struggle to differentiate between similar inputs that would
require different kernels. As such having the extra choice actually hinders the
model's performance rather than allowing it to achieve better performance.

When comparing the three different decision tree configurations, the performance
data does not support the theory that the tree may overfit to the training data.
The more limited trees (B and C) tend to perform worse than the unlimited
decision tree (A), though the numbers are not clear. When integrating the
decision tree into the SYCL library it is helpful to provide some limits, so as
to avoid heavily nested if statements and branching code.

%% file: vgg.tex
\section{Testing a full ML model}\label{sec:vgg}

This work was carried out to help provide general purpose compute libraries to
accelerate machine learning applications. Comparing the inference time of a
machine learning model using these techniques to similar libraries that use ore
manual tuning techniques can show the efficacy of this work.

One of the popular image classification models a few years ago was
VGG16~\cite{vgg}, developed at the Oxford Visual Geometry Group in 2015. By modern
standards it is a simple neural network made up of 16 convolutional and pooling
layers. Despite the small number of layers it has more parameters than most
modern networks with 138 million, as the convolutional layers have many
features.

While no longer state of the art, this model is still regularly used by machine
learning practitioners and much simpler than more recent image classification
networks, making it a good candidate to use to evaluate the performance of the
kernel selection process. Comparing the performance of individual kernels provides a
good proxy to determine how well a system will perform, but an evaluation on the
full system will help uncover any assumptions and shortcomings that would not be
visible at the micro-benchmark scale.

A SYCL-DNN sample implements the VGG16 network in SYCL using the pretrained
weights provided by the Keras Applications~\cite{keras_applications} Python
module. It can perform image classification based on the ImageNet dataset,
providing the class of an input image from the 1000 different ImageNet classes.
This pretrained network achieves 71.3\% performance classifying the top class of
an image in the ImageNet dataset. It is not the best performing model available
through Keras but is one of the simplest to implement.

In addition to testing the performance of this network on the devices discussed
earlier in this paper, we also tested two additional OpenCL devices. The kernels
used were tuned for each device using the methods discussed above and the
resulting deployment and selection algorithms were integrated into SYCL-DNN.

The devices used to test these techniques were:
\begin{itemize}
\item AMD R9 Nano GPU
\item Intel i7-6700K CPU
\item Intel HD 530 Gen9 GPU
\item ARM Mali G71 GPU
\end{itemize}

\begin{figure*}
  \centering
  \includegraphics{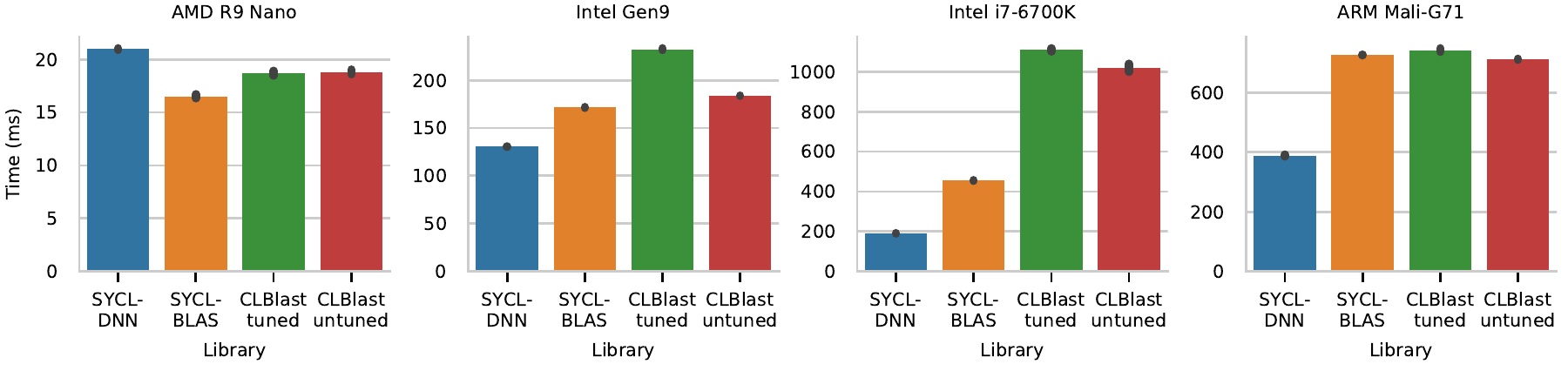}
\caption{The inference time in miliseconds of a single image using the VGG16
  model implemented using SYCL-DNN and different matrix backends when run on
  different devices.}
\label{fig:vgg_time}
\end{figure*}

\subsection{Library comparisons}

SYCL-DNN allows users to specify different backends that provide the matrix
multiplication routines used in neural networks. The library provides its own
matrix multiplication, but if a platform has access to a BLAS or other matrix
library then it can be easily integrated to make use of these optimized
routines. This functionality was used to provide comparisons to the tuned
SYCL-DNN matrix multiplication kernels, using both a SYCL-BLAS~\cite{syclblas}
backend and a CLBlast~\cite{nugteren2018clblast} backend.

SYCL-BLAS is another library developed by Codeplay to provide basic linear
algebra kernels. Designed with expression trees and templated kernels it allows
users to easily fuse kernels together at compile time, reducing the need to load
and store data between kernel launches, and is optimized for a range of devices.
SYCL-BLAS provides a number of different matrix multiplication routines,
including ones utilizing local (or shared) memory and ones designed for tall
skinny matrices that compute partial results which are combined in a final
reduction.  These kernels are significantly more sophisticated than the simple
kernel studied in this paper, however the parameters are all tuned by hand
requiring significant developer effort and time.

CLBlast is an OpenCL based BLAS library designed to be performant on a wide
range of OpenCL devices. It includes an automated tuning system to select the
optimal kernels for different devices, though this system is limited to
selecting the single best kernel for each device. Before running this benchmark,
the CLBlast library was tuned for each of the benchmark devices used. Similarly
to SYCL-BLAS, the CLBlast library contains multiple implementations of matrix
multiplication kernels to help achieve performance for different matrix shapes.

\subsection{Results}

The model was executed a number of times to accurately measure the time of
completion. A single image was used as an input, and the model classifies the
contents of that image. The weights and initial image are all transferred to the
compute device before starting timing, so the benchmark time only includes the
computation and not data transfer. The SYCL-DNN matrix multiplication routine
was tuned to use 8 kernel configurations per device selected using
PCA+$K$-means and a decision tree based runtime selection process. As discussed
in Sections~\ref{sec:pruning} and~\ref{sec:classification} these approaches give
good performance for different matrix sizes and devices.

Figure~\ref{fig:vgg_time} shows the execution time to compute one inference
using the VGG16 model. The different devices perform significantly differently
as would be expected as they have vastly different compute resources available.

The AMD R9 Nano performed an inference in less than 20ms using the optimized and
tuned matrix multiplication kernels from SYCL-BLAS and CLBlast.  This GPU along
with this particular machine learning model was one of the main targets of
optimization during the development of SYCL-BLAS so it is expected that it
performs well, outperforming both the kernel studied in this paper and CLBlast.
The SYCL-DNN kernel achieved times that were not far off the others, despite the
kernel being much simpler than those in the heavily optimized libraries and not
making use of the GPU's fast local memory.

By default CLBlast will use generic tuning parameters based on similar devices,
so for the R9 Nano the parameters are based on similar AMD cards. Tuning CLBlast
for this specific GPU using the provided tuning tools didn't provide any
benefit, though the actual kernels used did change. For the other devices the
tuning often had a negative impact on the performance of CLBlast. This is likely
to be a result of the limited way that the tuning works causing it to optimize
for best results on matrix sizes that differ from those used in the VGG16 model.
The GEMM routine in particular is tuned for single matrices of size 1024x1042
and 256x256, whereas the inputs to GEMM used in the model have a batch size of
16 and vary from 12544x64 to 512x512.

For the Intel CPU and integrated GPU the SYCL-DNN kernel actually performed
better than the optimized libraries. The CPU has very different performance
characteristics and compute resources to any of the GPUs, and CLBlast
particularly struggles to adapt to this.

Both SYCL-BLAS and CLBlast achieve similar performance on the ARM Mali GPU,
taking over 700ms per inference. SYCL-DNN on the other hand achieves under
400ms per inference, as it makes use of 4 different configurations out of the
chosen 8. This variety of possible kernel configurations allows the library to
hand the different matrix sizes where the other libraries only use a single
kernel configuration.

One of the areas where the SYCL-DNN kernels are at a disadvantage to the other
libraries is in the final fully connected layers in the model. These fully
connected layers are implemented as a matrix multiplication, but when using a
single image the activation tensor is actually a one dimensional vector rather
than a matrix. As such it is much more efficient to use a dedicated
matrix-vector multiplication routine common in BLAS libraries. The SYCL-DNN
kernel is comparatively inefficient in this case, as it is designed to compute
2D tiles of the output, which would only be one-dimensional. Despite this, the
library manages to provide sufficient performance on these operations that
the automatically tuned SYCL-DNN kernels outperform the other libraries overall.

%% file: conclusions.tex
\section{Conclusions}

Auto-tuning allows libraries to achieve performance on a wide range of
devices without requiring vast amounts of developer effort to adapt kernels and
routines to new hardware. In this paper we used a matrix multiplication case
study to evaluate some methods to allow auto-tuning to be deployed in compiled
SYCL libraries, balancing binary size, performance and adaptability to unseen
inputs.

Unsupervised machine learning techniques like clustering provide effective
methods to reduce the large kernel parameter space for a wide range of different
input sizes without sacrificing much performance. Some of these methods proved
more reliable and resilient than others, with some of the more advanced methods
like density based clustering methods struggling to provide performant kernels
in some cases.

One of the concerns raised in the original paper~\cite{ashes_paper} introducing
these ideas was that the techniques may rely too heavily on the dense benchmark
timing information. Intelligent auto-tuning techniques only sample from the very
large kernel parameter space, while the data collected for this study used a
comparatively small parameter space and so used a brute-force benchmarking
technique. The normalization techniques discussed in Section~\ref{sec:norm}
introduce sparsity into the data and Section~\ref{sec:pruning} shows that while
this does have an impact on the performance of the kernel selection routines,
this difference is minimal. This is promising for extending these results to
more complicated kernels that use more parameters that can take a larger range
of values.

After selecting the kernels to deploy in the SYCL library, there needs to be a
runtime routine to choose which of these kernels to execute for any given input.
The techniques discussed in Section~\ref{sec:classification}
show that decision trees can provide good performance, as well as
being easy to implement and integrate into a library.

When integrated into SYCL-DNN, these techniques met or vastly exceeded other
optimized BLAS libraries for a representative machine learning model. The
performance was competitive on a range of devices, from powerful desktop GPUs
through to embedded mobile GPUs, even though the kernels themselves are
relatively simple and don't use as many hardware features as those in the other
libraries.

Overall these tuning and deployment techniques provide an
efficient subset of all possible kernels where the kernels have to be provided
in binary format as with SYCL. These completely automated approaches allow
new devices to be supported with very little developer effort and relatively
small code changes.